\documentclass[12pt]{article}
\usepackage{epsfig}
\usepackage{pst-plot,url,epsf}
\usepackage{axodraw}
\newcommand{\AmS}{{\protect\the\textfont2
   A\kern-.1667em\lower.5ex\hbox{M}\kern-.125emS}}

\newcommand{\beq}{\begin{equation}}
\newcommand{\eeq}{\end{equation}}
\newcommand{\bea}{\begin{eqnarray}}
\newcommand{\eea}{\end{eqnarray}}
\newcommand{\beas}{\begin{eqnarray*}}
\newcommand{\eeas}{\end{eqnarray*}}
\newcommand{\nobody}{\rule{0ex}{1ex}}

\newcommand{\epm}{e^+e^-}

\newcommand{\ra}{\rightarrow}

\newcommand{\eeudmn}{e^+ e^- \ra u \bar{d} \mu^- \bar{\nu}_{\mu}}

\newcommand{\eemmbb}{e^+ e^- \ra \mu^+ \mu^- \bar{b} b}
\newcommand{\nn}{\nonumber}

\begin{document}
\begin{flushright}
DESY-05-046\\
SFB/CPP-05-08\\
2$^{\rm nd}$ revised version\\
\end{flushright}

\begin{center}
{\LARGE\bf One-loop electroweak factorizable corrections\\[2mm]
            for the Higgsstrahlung at a linear collider\footnote{Work supported
            in part by the Polish State Committee for Scientific Research
            in years 2004--2005 as a research grant, by the European
            Community's Human Potential Program under contracts
            HPRN-CT-2000-00149 Physics at Colliders and CT-2002-00311
            EURIDICE, and by DFG under Contract SFB/TR 9-03.}}

\vspace*{2cm}
Fred Jegerlehner$^{\rm a,}$\footnote{E-mails: fred.jegerlehner@desy.de,
kolodzie@us.edu.pl, twest@server.phys.us.edu.pl}
Karol Ko\l odziej$^{\rm b, 2}$
and Tomasz Westwa\'nski$^{\rm b, 2}$
\vspace{0.5cm}\\
$\nobody^{\rm a}${\small\it Deutsches Elektronen-Synchrotron DESY,
Platanenallee 6, D-15738 Zeuthen, Germany}\\
$\nobody^{\rm b}${\small\it
Institute of Physics, University of Silesia, ul. Uniwersytecka 4,
PL-40007 Katowice, Poland}\\
\vspace*{4.5cm}
{\bf Abstract}\\
\end{center}
We present standard model predictions for the four-fermion reaction $\eemmbb$
being one of the best detection channels of a low mass Higgs boson
produced through the Higgsstrahlung mechanism at a linear collider.
We include leading virtual and real QED corrections due to initial state
radiation and a modification of the Higgs--$b\bar b$ Yukawa
coupling, caused by the running of the $b$-quark mass, for $\eemmbb$.
The complete $\mathcal{O}(\alpha)$
electroweak corrections to the $Z$--Higgs production and to the $Z$
boson decay width, as well as remaining QCD and EW corrections to
the Higgs decay width, as can be calculated with a program {\tt HDECAY},
are taken into account in the double pole approximation.

\vfill

\newpage

\section{INTRODUCTION}

Although the Higgs boson has not yet been discovered,
its mass $m_H$ can be constrained in the framework of the standard model (SM)
by the virtual effects it has on precision electroweak (EW) observables.
Recent global fits to all precision EW data \cite{Hmass} give a central value
of $m_H=114_{-40}^{+56}$ GeV and an upper limit of 241~GeV, both at
95\% CL, in agreement with
combined results on the direct searches for the Higgs boson at LEP that
lead to a lower limit of $114.4$~GeV at 95\% CL \cite{LEPdir}.
These constraints indicate the mass range where the SM Higgs boson should be
searched for.
If the Higgs boson exists, it is most probably to be discovered
at the Large Hadron Collider, but its properties can be best investigated
in a clean experimental environment of $\epm$ collisions at a future
International Linear Collider (ILC) \cite{ILC}.

\begin{figure}[htb]
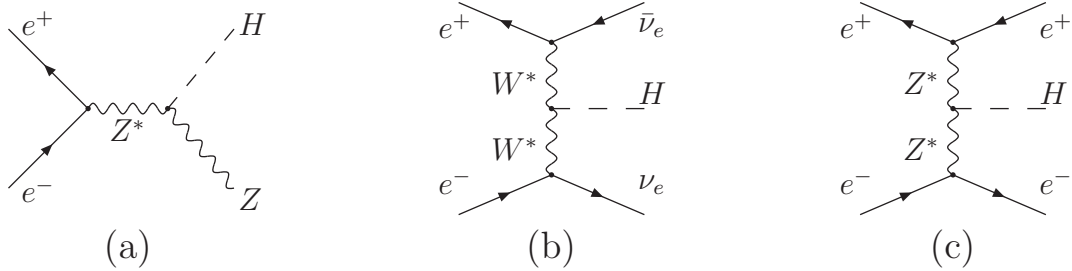

\vspace{120pt}
\centerline{
\includegraphics{fig1a.epsi}
\hfill
\hspace{0.5cm}
\includegraphics{fig1b.epsi}
\hfill
\includegraphics{fig1c.epsi}
\hfill
}
\caption{Feynman diagrams of reactions (\ref{eeZH}), (\ref{WW}) and
          (\ref{ZZ}), respectively.}
\label{fig:diag0s}
\end{figure}

Main mechanisms of the SM Higgs boson production at the ILC
are the Higgsstrahlung reaction
\bea
\label{eeZH}
                      e^+e^- \rightarrow  Z H,
\eea
the $WW$ fusion
\bea
\label{WW}
           e^+e^- \ra \nu_e\bar{\nu}_e W^*W^* \ra  \nu_e\bar{\nu}_e H,
\eea
and the $ZZ$ fusion process
\bea
\label{ZZ}
  e^+e^- \ra \epm Z^*Z^* \ra  \epm H.
\eea
The Feynman diagrams of reactions (\ref{eeZH}), (\ref{WW}) and (\ref{ZZ})
are depicted in Fig.~\ref{fig:diag0s}a, ~\ref{fig:diag0s}b and
~\ref{fig:diag0s}c, respectively.

The cross section of reaction (\ref{eeZH}) decreases according to
the $1/s$ scaling
law, while that of reaction (\ref{WW}) grows as $\ln(s/m_H^2)$. Hence, while
the Higgsstrahlung dominates the Higgs boson production at low energies,
the $WW$ fusion process overtakes it at higher energies. The Higgs boson
production rate through the $ZZ$ fusion mechanism (\ref{ZZ})
  is by an order of magnitude smaller than that of process
(\ref{WW}). The production of the SM Higgs boson in the intermediate
mass range at high-energy $\epm$
colliders through reaction (\ref{eeZH}) was already thoroughly studied
in the literature \cite{inthiggs}. In the present paper, we contribute
further to the theoretical analysis of the Higgsstrahlung
reaction by taking into account decays of the $Z$ and Higgs bosons,
including the most relevant EW radiative effects both to the production
and decay subprocesses. Preliminary results of the present study
have been already presented in our conference paper \cite{JKW}.

\begin{figure}[htb]
\vspace{140pt}
\includegraphics{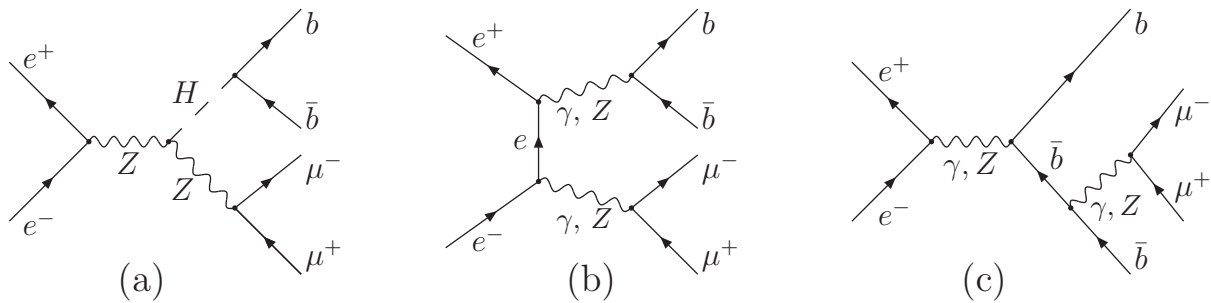}
\caption{Examples of Feynman diagrams of reaction (\ref{bmmb}):
(a) the double resonance `signal', (b) and (c) `background'
diagrams.}
\label{fig:diags}
\end{figure}

If the Higgs boson has mass in the lower part of the range indicated above,
say $m_H < 140$~GeV, it would decay dominantly into a
$\bar{b} b$ quark pair. As the $Z$ boson of reaction (\ref{eeZH}) decays into
a fermion--antifermion pair too, one actually observes the Higgsstrahlung
through reactions with four fermions in the final state.
To be more specific, in the following we will concentrate on a
four-fermion channel
\bea
\label{bmmb}
  e^+(p_1)+e^-(p_2) \rightarrow \mu^+(p_3)+ \mu^-(p_4)
                                             +\bar{b}(p_5)+ b(p_6),
\eea
one of the most relevant for detection of (\ref{eeZH}) at the ILC,
and the corresponding bremsstrahlung reaction
\bea
\label{bmmbg}
  e^+(p_1)+e^-(p_2) \rightarrow \mu^+(p_3)+ \mu^-(p_4)
                    +\bar{b}(p_5)+ b(p_6) + \gamma(p_7).
\eea
Both in (\ref{bmmb}) and (\ref{bmmbg}), the particle four momenta
have been indicated in parentheses.
While final states of reactions (\ref{bmmb}) and (\ref{bmmbg}) are 
practically not
detectable in hadronic collisions because
of the overwhelming QCD background, at the ILC, they
should leave a particularly clear signature in a detector.
To the lowest order of the SM, in the unitary gauge and neglecting the Higgs
boson
coupling to the electron, reactions (\ref{bmmb}) and (\ref{bmmbg}) receive
contributions from 34 and 236 Feynman diagrams, respectively. Typical
examples of Feynman diagrams of reaction (\ref{bmmb})  are
depicted in Fig.~\ref{fig:diags}. The Higgsstrahlung `signal' diagram is
shown in Fig.~1a,
while the diagrams in Figs.~1b and 1c represent typical `background'
diagrams. The Feynman diagrams of reaction (\ref{bmmbg}) are obtained
from those of reaction (\ref{bmmb}) by attaching an external photon line
to each electrically charged particle line. Reactions
(\ref{bmmb}) and (\ref{bmmbg}) are typical examples of the
neutral current four-fermion reactions and the corresponding hard
bremsstrahlung reactions which had been already studied
for LEP2 and TESLA TDR \cite{ILC} in the case of massless fermions
\cite{EXCALIBUR}, \cite{Racoon} and without neglecting fermion masses
\cite{all4f}, \cite{Krauss}. Reaction (\ref{bmmb}) was
in particular studied in \cite{Boos}. It was also considered, among other
four-fermion reactions relevant for the Higgs boson production and decay,
in a study of the Higgs boson searches at LEP2 performed in \cite{Passarino}, 
where references to other works dedicated to the subject can be found, too.

In order to match the precision of data from a high luminosity
linear collider, expected to be better than 1\%, it is necessary to include
radiative corrections in the SM predictions for reaction (\ref{bmmb}).
Calculation of the complete electroweak (EW) $\mathcal{O}(\alpha)$
corrections to a four-fermion reaction like (\ref{bmmb}) is a
challenging task. Problems encountered in the first attempt
of such a complete calculation were described in \cite{Vicini}.
Substantial progress in a full one-loop calculation for
$\eeudmn$ was reported by the GRACE/1-LOOP team \cite{GRACE} and
recently results of the first calculation of the complete EW
$\mathcal{O}(\alpha)$ corrections to charged-current
$\epm \ra 4$~fermion processes has been presented \cite{DDRW}.
Another step towards obtaining high precision predictions for the Higgs boson
production, accomplished in \cite{nnH}, was calculations of EW corrections to
$e^+ e^- \ra \nu \bar{\nu} H$, a process related to (\ref{bmmb}).
At the moment, however, there is no complete calculation of the corrections
to neutral-current $\epm \ra 4$~fermion processes available that could
be included into a Monte Carlo (MC) generator. Therefore it
seems natural to include the leading QED corrections
and to apply the double pole approximation (DPA) for the
EW $\mathcal{O}(\alpha)$ corrections, as it had been done before
in \cite{DPAWW} in the case of $W$-pair production  at LEP2.
This means that we will include the so called factorizable EW corrections
to a $ZH$ production (\ref{eeZH}) and to subprocesses of the $Z$ and Higgs
decay, which are available in the literature. 

To the lowest order of the SM, the cross sections of reactions (\ref{bmmb})
and (\ref{bmmbg}) can be computed with a program {\tt ee4fg} \cite{ee4fg}.
On the basis of {\tt ee4fg}, we have written a dedicated program
{\tt eezh4f} that includes the following  
radiative corrections to (\ref{bmmb}):
\begin{itemize}
\item the virtual and real soft photon QED corrections to the
       on-shell $ZH$ production (\ref{eeZH}), a universal part of which
       is utilized for all Feynman diagrams of  reaction (\ref{bmmb})
       and combined with the initial state hard bremsstrahlung correction
       of (\ref{bmmbg}), which we refer to as the initial state radiation
       (ISR) QED correction;
\item the weak corrections to the $Z$--Higgs production
        \cite{FJ1}, \cite{ewZH}, {\em i.e.} the
       infrared (IR) finite part of the complete $\mathcal{O}(\alpha)$ EW
       correction to (\ref{eeZH}) and
       full EW corrections, including hard photon emission, to the $Z$ decay
       width \cite{ewZ}, \cite{FJ2}  in the DPA;
\item the correction Higgs boson decay width, as can be computed with
       {\tt HDECAY} \cite{HDECAY}.
\end{itemize}
The correction to the partial Higgs boson decay width $\Gamma_{H\ra b\bar b}$
is split into two parts: one part is the tree level partial width
$\Gamma^{(0)}_{H\ra b\bar b}$ parametrized in terms of the running
$b$-quark mass and the other part is the correction term that contains
the remaining QCD corrections and the bulk of EW corrections, as described
in \cite{HDECAY}. The first part is included in the calculation of
reactions (\ref{bmmb}) and (\ref{bmmbg}) through the corresponding
modification of the $H\ra b\bar b$ Yukawa coupling, while the second part
is taken into account in the DPA.

The non-factorizable corrections to (\ref{bmmb}) have not been included
in {\tt eezh4f} yet. 
Although, a comparison with the related process of $ZZ$ production and 
decay, where the non-factorizable corrections largely cancel each other in the 
total cross section \cite{BCB}, \cite{DD}, may suggest that
they are also numerically suppressed for the $ZH$-mediated four-fermion final
states if decay angles are integrated over,
they should be included in the program if one wants
to study differential cross sections, or if one wants to improve predictions
for invariant-mass distributions and similar quantities.

\section{SCHEME OF THE CALCULATION}

The corrections listed in the introduction are taken into account in the
radiatively corrected total cross section of (\ref{bmmb}) according to
the master formula
\bea
\label{totalcs}
\int{\rm d}\sigma = \int{\rm d}\sigma_{\rm Born+r.m.} +
\int_{E_{\gamma}<E_{\rm cut}}
                 {\rm d}\sigma_{\rm virt + soft, univ.}^{\rm QED\;ISR}
+\int_{E_{\gamma}>E_{\rm cut}}{\rm d}\sigma_{\rm hard}^{\rm QED\;ISR}
+\int{\rm d}\sigma_{\rm virt, finite}^{\rm EW\;DPA}.
\eea
By ${\rm d}\sigma_{\rm Born+r.m.}$ we denote the effective Born
approximation
\bea
\label{Born}
{\rm d}\sigma_{\rm Born+r.m.}
=\frac{1}{2s}\left\{\left|M_{\rm Born}^{4f}\right|^2\;
+ 2\;{\rm Re}\left(M_{\rm Born}^{4f\;^*}
\;\delta M_{\rm r.m.}^{4f}\right)\right\}{\rm d}\Phi_{4f},
\eea
where $M_{\rm Born}^{4f}$ is the matrix element of reaction (\ref{bmmb})
obtained with the complete set of the lowest order Feynman diagrams,
$\delta M_{\rm r.m.}^{4f}$ is the correction to the amplitude
of the `signal' diagram of Fig.~\ref{fig:diags}(a) due to
modification $\delta g_{Hb\bar b}^{\rm r.m.}$ of the lowest order
Higgs--$b\bar b$ Yukawa coupling $g_{Hb\bar b}$
caused by the running of the $b$-quark mass,
${\rm d}\Phi_{4f}$ is the Lorentz invariant four-particle phase
space element and $4f$ is a shorthand notation for the final state
of reaction (\ref{bmmb}). The correction $\delta M_{\rm r.m.}^{4f}$ can be
written as
\bea
\label{deltam}
\delta M_{\rm r.m.}^{4f}&=&\bar{v}\left(p_1\right)
\left(A_1^{(0)}\gamma_{\mu} + A_2^{(0)}\gamma_{\mu}\gamma_5\right)
            u\left(p_2\right) \frac{-g^{\mu\nu}
+p_{34}^{\mu}p_{34}^{\nu}/M_Z^2}{D_Z(p_{34})D_H(p_{56})}\nn\\
&\times&\bar{u}\left(p_4\right)\left(a_1\gamma_{\nu}
   +a_2\gamma_{\nu}\gamma_5\right)v\left(p_3\right)
\; \delta g_{Hb\bar b}^{\rm r.m.}\;\bar{u}(p_6)v(p_5),
\eea
where $A_1^{(0)}$ and $A_2^{(0)}$ are the lowest order invariant amplitudes
of the on-shell Higgsstrahlung reaction (\ref{eeZH}) \cite{FJ1} which read
\bea
\label{ai0}
A_i^{(0)}=\frac{g_{HZZ}^{(0)}}{s-M_Z^2}\;{a_i^{(0)}},\qquad i=1,2
\eea
and $D_Z(p_{34})$ and $D_H(p_{56})$ are
denominators of the $Z$ and Higgs boson propagators
\bea
\label{props}
D_Z(p_{34})={\left(p_3+p_4\right)^2-M_Z^2}, \qquad
D_H(p_{56})={\left(p_5+p_6\right)^2-M_H^2},
\eea
with the complex mass parameters
\bea
\label{cmass}
M_Z^2=m_Z^2-i m_Z \Gamma_Z, \qquad M_H^2=m_H^2-i m_H \Gamma_H.
\eea
The `fixed' total widths $\Gamma_Z$ and $\Gamma_H$ of the $Z$ and Higgs
bosons have been introduced in order to avoid singularities in resonant
regions of the corresponding propagators.
We have used the complex mass substitution also in Eq.~(\ref{ai0}),
although the propagator factor of Eq.~(\ref{ai0}) can never become
resonant as we want to keep open a possibility of performing
calculations in the complex mass scheme \cite{Racoon}, which preserves
Ward identities.

To lowest order of the SM, the vector and axial-vector couplings of the
$Z$ boson to leptons, $a_1^{(0)}$, $a_2^{(0)}$, the Higgs boson coupling
to $Z$, $g_{HZZ}^{(0)}$ of Eq.~(\ref{ai0}) and the Higgs--$b \bar b$
Yukawa coupling, $g_{Hbb}^{(0)}$, are given by
\bea
\label{Zll}
a_1^{(0)}=\frac{4s_W^2-1}{4s_Wc_W}\;e_W,
\qquad a_2^{(0)}=\frac{e_W}{4s_Wc_W},
\qquad g_{HZZ}^{(0)}=\frac{e_Wm_Z}{s_Wc_W},
\qquad g_{Hbb}^{(0)}=-\frac{e_Wm_b}{2s_Wm_W},
\eea
with the effective electric charge $e_W=\left(4\pi\alpha_W\right)^{1/2}$
and electroweak mixing angle $\theta_W$ defined by
\beq
\label{awsw}
\alpha_W=\sqrt{2} G_{\mu} m_W^2 s_W^2/\pi, \qquad {\rm and} \qquad
                 s_W^2=1-m_W^2/m_Z^2,
\eeq
where $m_W$ and $m_Z$ are the physical masses of the $W$ and $Z$ boson,
respectively. This choice, which is exactly equivalent
to the $G_{\mu}$-scheme of \cite{FJ1}, is in the program referred
to as the {\em `fixed width scheme'} (FWS).
We have introduced usual shorthand notation $s_W=\sin\theta_W$ and
$c_W=\cos\theta_W$ in Eqs.~(\ref{Zll}) and (\ref{awsw}).

The total widths of Eq.~(\ref{cmass}) are calculated numerically in the
framework of the SM: $\Gamma_Z$ is computed
with a program based on reference \cite{FJ2} and $\Gamma_H$ is obtained
with a program {\tt HDECAY} \cite{HDECAY}, both programs including
radiative corrections.
In order not to violate unitarity it is important that we include
exactly the same corrections in the amplitudes of the partial
decay widths $\Gamma_{Z \ra \mu^+\mu^-}$ and $\Gamma_{H \ra b\bar b}$.

The program {\tt HDECAY} includes the full massive NLO QCD corrections
close to the thresholds and the massless $\mathcal{O}(\alpha^3_s)$
corrections far above the thresholds. Both regions are related
with a simple linear interpolation equation which, for the partial
Higgs decay width $\Gamma_{H \ra b\bar b}$ reads
\bea
\label{interpol}
\Gamma_{H \ra b\bar b}=
\left(1-r^2\right) \Gamma_{H \ra b\bar b}\Big(\overline{m}_b(m_H)\Big)
+ r^2 \Gamma_{H \ra b\bar b}\left(m_b\right),
\qquad {\rm with} \quad r=\frac{2m_b}{m_H},
\eea
where large logarithms are resummed in the running $b$-quark quark mass
in the $\overline{\rm MS}$ renormalization scheme $\overline{m}_b(m_H)$.

By inserting the representations
\bea
\Gamma_{H \ra b\bar b} \Big(\overline{m}_b(m_H)\Big) &=&
\Gamma_{H \ra b\bar b}^{(0)}\Big(\overline{m}_b(m_H)\Big) +
              \Delta \Gamma_{H \ra b\bar b}\Big(\overline{m}_b(m_H)\Big),\\
\Gamma_{H \ra b\bar b}\left(m_b\right)&=&
\Gamma_{H \ra b\bar b}^{(0)}\left(m_b\right) +
              \Delta \Gamma_{H \ra b\bar b}\left(m_b\right),
\eea
with the lowest order partial width into $b\bar b$-quark-pair given by
\bea
\Gamma_{H \ra b\bar b}^{(0)}\left(m_b\right)=
3\frac{{g_{Hb\bar b}^{(0)}}^2}{8\pi}m_H\left(1
-\frac{4m_b^2}{m_H^2}\right)^{\frac{3}{2}},
\eea
into Eq.~(\ref{interpol}), we obtain the representation
\bea
\label{gamma}
\Gamma_{H \ra b\bar b}=\Gamma_{H \ra b\bar b}^{\rm r.m.}+
\Delta \Gamma_{H \ra b\bar b}
\eea
for $\Gamma_{H \ra b\bar b}$.
The first term on the right hand side of Eq.~(\ref{gamma})
\bea
\label{gamrm}
\Gamma_{H \ra b\bar b}^{\rm r.m.}=
\left(1-r^2\right) \Gamma_{H \ra b\bar b}^{(0)}\Big(\overline{m}_b(m_H)\Big)
+ r^2 \Gamma_{H \ra b\bar b}^{(0)}\left(m_b\right)
\eea
contains the correction due to running of the $b$-quark mass, while
the second term, $\Delta \Gamma_{H \ra b\bar b}$,
incorporates the remaining QCD corrections and
EW corrections that are taken into account in {\tt HDECAY}.

Now, if we write the radiatively corrected Higgs--$b\bar b$ Yukawa coupling
in the form
\bea
g_{Hb\bar b}=g_{Hb\bar b}^{(0)}+\delta g_{Hb\bar b}^{\rm r.m.}
+\delta g_{Hb\bar b}
\eea
and the corrected partial Higgs decay width as
\bea
\Gamma_{H \ra b\bar b}&=&\Gamma_{H \ra b\bar b}^{(0)}(m_b)\left[1
+ 2\;{\rm Re}{\left(\frac{\delta g_{Hb\bar b}^{\rm r.m.}
+\delta g_{Hb\bar b}}{g_{Hb\bar b}^{(0)}}\right)} \right],
\eea
then we will obtain the following expressions for the radiative
corrections $\delta g_{Hb\bar b}^{\rm r.m.}$ and $\delta g_{Hb\bar b}$,
assuming that they real,
\bea
\delta g_{Hb\bar b}^{\rm r.m.}=\frac{\Gamma_{H \ra b\bar b}^{\rm r.m.}
-\Gamma_{H \ra b\bar b}^{(0)}(m_b)}{2\Gamma_{H \ra b\bar b}^{(0)}(m_b)}\;
g_{Hb\bar b}^{(0)},
\qquad
\delta g_{Hb\bar b}=\frac{\Gamma_{H \ra b\bar b}
-\Gamma_{H \ra b\bar b}^{\rm r.m.}}{2\Gamma_{H \ra b\bar b}^{(0)}(m_b)}\;
g_{Hb\bar b}^{(0)}.
\eea
We see that taking into account the correction
$\delta g_{Hb\bar b}^{\rm r.m.}$ in Eq.~(\ref{Born}) is actually
equivalent to replacing $m_b$ in the lowest
order Higgs--$b\bar b$ Yukawa coupling of Eq.~(\ref{Zll})
by some effective value of the $b$-quark mass. The same modification of
the Higgs--$b\bar b$ Yukawa coupling is done in the soft and hard
bremsstrahlung corrections represented by the second and third term
on the right hand side of Eq.~(\ref{totalcs}), respectively.

After having arranged the radiative corrections in the manner just described,
the soft brems\-strah\-lung contribution of Eq.~(\ref{totalcs})
can be written as
\bea
\label{QEDvs}
{\rm d}\sigma_{\rm virt + soft, univ.}^{\rm QED\;ISR}=
{\rm d}\sigma_{\rm Born+r.m.}C_{\rm virt + soft, univ.}^{\rm QED\;ISR},
\eea
where the correction factor $C_{\rm virt + soft, univ.}^{\rm QED\;ISR}$,
\bea
C_{\rm virt + soft, univ.}^{\rm QED\;ISR}=
\frac{e^2}{2\pi^2} & &\hspace*{-0.5cm}\left[\left(\ln\frac{s}{m_e^2}-1\right)
          \ln\frac{2E_{\rm cut}}{\sqrt{s}}
         +\frac{3}{4}\ln\frac{s}{m_e^2} \right],
\label{CQEDvs}
\eea
combines the universal IR singular part of the
$\mathcal{O}(\alpha)$ virtual QED correction to the on-shell $ZH$ production
process (\ref{eeZH}) with the soft bremsstrahlung correction to
(\ref{bmmb}), integrated up to the soft photon energy cut $E_{\rm cut}$.
In Eq.~(\ref{QEDvs}), $e$ is the electric charge that is given in terms
of the fine structure constant in the Thomson limit $\alpha_0$,
$e=\left(4\pi\alpha_0\right)^{1/2}$.

The third term on the right hand side of Eq.~(\ref{totalcs}) represents
the initial state real hard photon correction to reaction (\ref{bmmb}),
{\em i.e.} the lowest order cross section of reaction
(\ref{bmmbg}) with the photon energy cut $E_{\gamma} > E_{\rm cut}$,
which is calculated taking into account the photon emission from
the initial state particles.
It has been checked numerically that the dependence on $E_{\rm cut}$
cancels in the sum of the second and third term on the right hand side
of Eq.~(\ref{totalcs}), provided that the real photon coupling
to the initial state fermions is parametrized in terms of $\alpha_0$, too.

Finally, the last integrand on the right hand side of Eq.~(\ref{totalcs}),
${\rm d}\sigma_{\rm virt, finite}^{\rm EW,\;DPA}$, is
the IR finite part of the virtual EW $\mathcal{O}(\alpha)$ correction
to reaction (\ref{bmmb}) in the DPA. It can be written in the following way
\bea
\label{DPA}
{\rm d}\sigma_{\rm virt, finite}^{\rm EW,\;DPA}=
\frac{1}{2s}\left\{
\left|{M^{(0)}_{DPA}}\right|^2 {C_{\rm QED}^{\rm non-univ.}}
+ 2{\rm Re}\left({M^{(0)^*}_{DPA}\delta M_{DPA}}\right)\right\}
{\rm d}\Phi_{4f},
\eea
where the lowest order matrix element $M^{(0)}_{DPA}$ and
the one-loop correction $\delta M_{DPA}$ in the DPA, which are given below,
are calculated with  the projected four momenta $k_i$, $i=3,...,6$, of
the final state particles, except for denominators of the $Z$ and Higgs
boson propagators. The projected four momenta are obtained from
the four momenta $p_i$, $i=3,...,6$, of reaction (\ref{bmmb}) with a, to some
extent arbitrary, projection procedure which will be defined later;
$C_{\rm QED}^{\rm non-univ.}$ denotes the IR finite
non universal constant part of the $\mathcal{O}(\alpha)$ QED correction
that has not been taken into account in Eq.~(\ref{CQEDvs}). It reads
\bea
\label{bremnu}
C^{\rm QED\; ISR}_{\rm non-univ.} =
    \frac{e^2}{2\pi^2} \left(\frac{\pi^2}{6}-1\right).
\eea

The matrix elements $M^{(0)}_{DPA}$ and $\delta M_{DPA}$
for the $Z$ and Higgs boson production and decay of Eq.~(\ref{DPA})
are given by
\bea
\label{m0}
M^{(0)}_{DPA}&=&-\bar{v}\left(p_1\right)
\left(A_1^{(0)}\gamma_{\mu} + A_2^{(0)}\gamma_{\mu}\gamma_5\right)
    u\left(p_2\right)\frac{1}{D_Z(p_{34})D_H(p_{56})}{N}_Z^{\mu} \; {N}_H,\\
\label{dm}
\delta M_{DPA}&=&-\left\{\bar{v}\left(p_1\right)\bigg[
\delta A_1\gamma_{\mu} + \delta A_2\gamma_{\mu}\gamma_5
+\left(A_{31}p_{1\mu} + A_{32}p_{2\mu}\right)/\!\!\!k_Z \right. \nn\\
& &\qquad\qquad\qquad\qquad\qquad\;\,+\left(A_{41}p_{1\mu}
+ A_{42}p_{2\mu}\right)
        /\!\!\!k_Z\gamma_5\bigg] u\left(p_2\right){N}_Z^{\mu} \; {N}_H \\
& &+\left. \bar{v}\left(p_1\right)\left(
A_1^{(0)}\gamma_{\mu} + A_2^{(0)}\gamma_{\mu}\gamma_5
\right) u\left(p_2\right)\bigg(\delta N_Z^{\mu} \; N_H
+ N_Z^{\mu} \; \delta N_H\bigg)\right\} \frac{1}{D_Z(p_{34})D_H(p_{56})}, \nn
\eea
with denominators of the $Z$ and Higgs boson propagators defined by
Eq.~(\ref{props}) and amplitudes of decay subprocesses given by
\bea
\label{n0}
{N}_Z^{\mu} =
\bar{u}\left(k_4\right)\left(a_1^{(0)}\gamma^{\mu}
+a_2^{(0)}\gamma^{\mu}\gamma_5\right) v\left(k_3\right), \qquad
{N}_H = g_{Hbb}^{(0)}\; \bar{u}\left(k_6\right)v\left(k_5\right)
\eea
in the lowest order and
\bea
\label{dn}
{\delta N}_Z^{\mu} =
\bar{u}\left(k_4\right)\left(\delta a_1\gamma^{\mu}
     +\delta a_2\gamma^{\mu}\gamma_5\right) v\left(k_3\right), \qquad
{\delta N}_H = \delta g_{Hbb}\; \bar{u}\left(k_6\right)v\left(k_5\right)
\eea
in the one-loop and higher order. Let us note that the invariant amplitudes
$\delta a_1$, $\delta a_2$ and $\delta g_{Hbb}$ contain the QED correction 
factors, which include the hard photon emission. We have neglected the 
so-called longitudinal part of the $Z$-boson propagator proportional to 
$p_{34}^{\mu}p_{34}^{\nu}$ in Eqs.~(\ref{m0})
and (\ref{dm}). Keeping this part would lead to some ambiguity, as
the propagator couples to the final state currents $N_Z^{\mu}$ 
and $\delta N_Z^{\mu}$ which are
parametized  in terms of the projected momenta $k_3$ and $k_4$. The neglected
term is of the order of $m_{\mu}$, which is consistent with neglect
of the $\mathcal{O}(m_{\mu})$ terms in calculation of the one-loop
EW decay amplitudes $\delta a_1$ and $\delta a_2$ of Eq.~(\ref{dn}).

The invariant amplitudes
of Eq.~(\ref{dm})\footnote{Only amplitudes which do not vanish in the limit
$m_e \ra 0$ are included.}
\bea
\label{ampZH}
\delta A_i=\delta A_i(s,\cos\theta), \qquad
A_{ji} = A_{ji}(s,\cos\theta), \quad i=1,2, \quad j=3,4,
\eea
represent the IR finite parts of the EW one-loop
correction to the on-shell $Z$--Higgs production process (\ref{eeZH}).
They are complex functions of
$s=(p_1+p_2)^2$ and $\cos\theta$, $\theta$ being the Higgs boson production
angle with respect to the initial positron beam in the centre of mass
system (CMS), computed with the program {\tt eezh4f}. The latter makes use
of the program worked out in
\cite{FJ1} and the package {\tt FF 2.0}. The package {\tt FF}
written by G.~J. van Oldenborgh allows to evaluate
one-loop integrals \cite{FF}. Note that we have changed a little the notation
in Eq.~(\ref{dm}) with respect to that of Eq.~(2.5) of \cite{FJ1}.
The IR finite contributions to the EW one-loop form factors of the
$Zll$-vertex $\delta a_1$, $\delta a_1$ and the correction
$\delta g_{Hbb}$ of the $Hb\bar b$-vertex
of Eq.~(\ref{dn}) are also calculated numerically
following references \cite{FJ2} and \cite{HDECAY}.

As the computation of the one-loop electroweak amplitudes of Eq.~(\ref{ampZH})
slows down the MC integration substantially, a simple
interpolation routine has been written that samples the amplitudes at
a few hundred values of $\cos\theta$
and then the amplitudes for all intermediate values of $\cos\theta$ are
obtained by a linear interpolation. This gives a tremendous gain in
the speed of computation, while there is practically
no difference between the results obtained with the interpolation
routine and without it.

We end this section with a description of the projection procedure we
have applied in the program.
The projected momenta $k_Z$ and $k_i$, of Eqs.~(\ref{dm}), (\ref{n0})
and (\ref{dn})
are obtained from the four momenta $p_i$, $i=3,...,6$, of the final state
fermions of reaction (\ref{bmmb}) with
the following projection procedure.

First the on-shell momenta and energies of the Higgs and $Z$ boson in the
CMS are fixed by
\parbox{8cm}{
\beas
\left|\vec{k}_{H}\right|&=&
\frac{\lambda^{\frac{1}{2}}\left(s,m_Z^2,m_H^2\right)}{2s^{\frac{1}{2}}},\\
E_{H}&=&\left(\vec{k}_{H}^2 + m_H^2\right)^{\frac{1}{2}},
\eeas}
\hfill
\parbox{6cm}{
\beas
\vec{k}_{H}&=&\left|\vec{k}_{H}\right|
\frac{\vec{p}_{5}+\vec{p}_{6}}
                  {\left|\vec{p}_{5}+\vec{p}_{6}\right|}, \\
\vec{k}_{Z}&=&-\vec{k}_{H}, \quad
E_{Z}=\sqrt{s}-E_{H}.
\eeas}
\hfill
\parbox{2cm}{\bea \label{momHZ}  \eea}


Next the four momenta $p_5$ and $p_3$ of reaction (\ref{bmmb})
are boosted to the rest frame of the $b\bar b$- and $\mu^+\mu^-$-pairs,
respectively, where they are denoted by $p'_5$ and $p'_3$. The projected four
momenta $k'_i$, $i=5,6$, in the rest frame of the $b\bar b$-pair
are then obtained by the kinematical relations\\
\parbox{8cm}{
\beas \qquad\qquad
\left|\vec{k'}_{5}\right|&=&
\frac{\lambda^{\frac{1}{2}}\left(m_H^2,m_5^2,m_6^2\right)}{2m_H}, \\
\vec{k'}_{6}&=&-\vec{k'}_{5},
\eeas}
\hfill
\parbox{6cm}{
\beas
\vec{k'}_{5}&=&
\left|\vec{k'}_{5}\right|\frac{\vec{p'}_5}{\left|\vec{p'}_{5}\right|},\\
{E'}_{i}&=&\left(\vec{k'}_{i}^2 + m_i^2\right)^{\frac{1}{2}}, \quad i=5,6.
\eeas}
\hfill
\parbox{2cm}{\bea \label{momdH} \eea}
Similarly, one obtains $k'_3$ and $k'_4$ in the rest frame of the
$\mu^+\mu^-$-pair using\\
\parbox{8cm}{
\beas \qquad\qquad
\left|\vec{k'}_{3}\right|&=&
\frac{\lambda^{\frac{1}{2}}\left(m_Z^2,m_3^2,m_4^2\right)}{2m_Z}, \\
\vec{k'}_{4}&=&-\vec{k'}_{3},
\eeas}
\hfill
\parbox{6cm}{
\beas
\vec{k'}_{3}&=&
\left|\vec{k'}_{3}\right|\frac{\vec{p'}_3}{\left|\vec{p'}_{3}\right|},\\
{E'}_{i}&=&\left(\vec{k'}_{i}^2 + m_i^2\right)^{\frac{1}{2}}, \quad i=3,4.
\eeas}
\hfill
\parbox{2cm}{\bea \label{momdZ} \eea}

The four momenta $k'_i$, $i=3,...,6$, are then boosted back
to the CMS giving the projected four momenta $k_i$,
$i=3,...,6$ of $\mu^+, \mu^-, \bar{b}$ and $b$ that satisfy
the necessary on-shell relations
\bea
k_3^2=k_4^2\!\!\!&=&\!\!\!m_{\mu}^2, \qquad k_5^2=k_6^2=m_{b}^2,\nn\\
\left(k_3+k_4\right)^2\!\!\!&=&\!\!\!m_Z^2,
\qquad \left(k_5+k_6\right)^2=m_H^2.\nn
\eea
The actual value of $\cos\theta$ in Eq.~(\ref{ampZH}) is given by
$\cos\theta=k_H^3/|\vec{k}_H|$.
The described projection procedure is not unique. As the Higgs boson width
is small,
the ambiguity between different
possible projections is mainly related to the off-shellness of the $Z$
boson and is of the order of $\alpha\Gamma_Z/(\pi m_Z)$.

\section{NUMERICAL RESULTS}

In this section, we will present a sample of numerical results for
reaction (\ref{bmmb}) which have been obtained with the current version of
a program {\tt eezh4f}. Computations have been performed
in the FWS with the $Z$ boson mass, Fermi
coupling and fine structure constant in the Thomson limit
as the initial
SM EW physical parameters \cite{PDG}:
\bea
\label{params1}
m_Z=91.1876\; {\rm GeV},\qquad
G_{\mu}=1.16639 \times 10^{-5}\;{\rm GeV}^{-2}, \qquad
\alpha_0=1/137.03599976.
\eea
The external particle masses of reaction (\ref{bmmb}) are the following:
\bea
\label{params3}
m_e=0.510998902\;{\rm MeV},\qquad m_{\mu}=105.658357\;{\rm MeV},
\qquad m_b=4.4\;{\rm GeV}.
\eea
For definiteness, we give also values of other fermion masses used in the
computation:\\
\parbox{5.5cm}{
\bea
\label{params4}
m_{\tau}&\!=\!&1.77699\;{\rm GeV},\nn\\
m_u&\!=\!&62\;{\rm MeV},\nn\eea}
\hfill
\parbox{5.5cm}{\beas
    m_c&\!=\!&1.5\;{\rm GeV},\\
    m_d&\!=\!&83\;{\rm MeV},\eeas}
\hfill
\parbox{5.5cm}{\bea
    m_t&\!=\!&177.7\;{\rm GeV},\nn\\
    m_s&\!=\!&215\;{\rm MeV}.\eea}

The light quark masses of Eq.~(\ref{params4}),
together with $\alpha_s=0.123$, reproduce
the hadronic contribution to the running of the fine structure constant.

Assuming a specific value of the Higgs boson mass, the $W$ boson mass and
the  total $Z$ boson width are calculated in a subroutine based on
\cite{FJ2}, while the total Higgs boson width is calculated with
{\tt HDECAY} \cite{HDECAY}.
We obtain the following values for them for $m_H=115$~GeV and parameters
specified in
Eqs.~(\ref{params1}--\ref{params4})\\
\bea
\label{params5}
m_W = 80.40844\; {\rm GeV}, \qquad \Gamma_Z = 2.50393\; {\rm GeV}
\qquad \Gamma_H=2.8542\; {\rm MeV}.
\eea

\begin{figure}[htb]
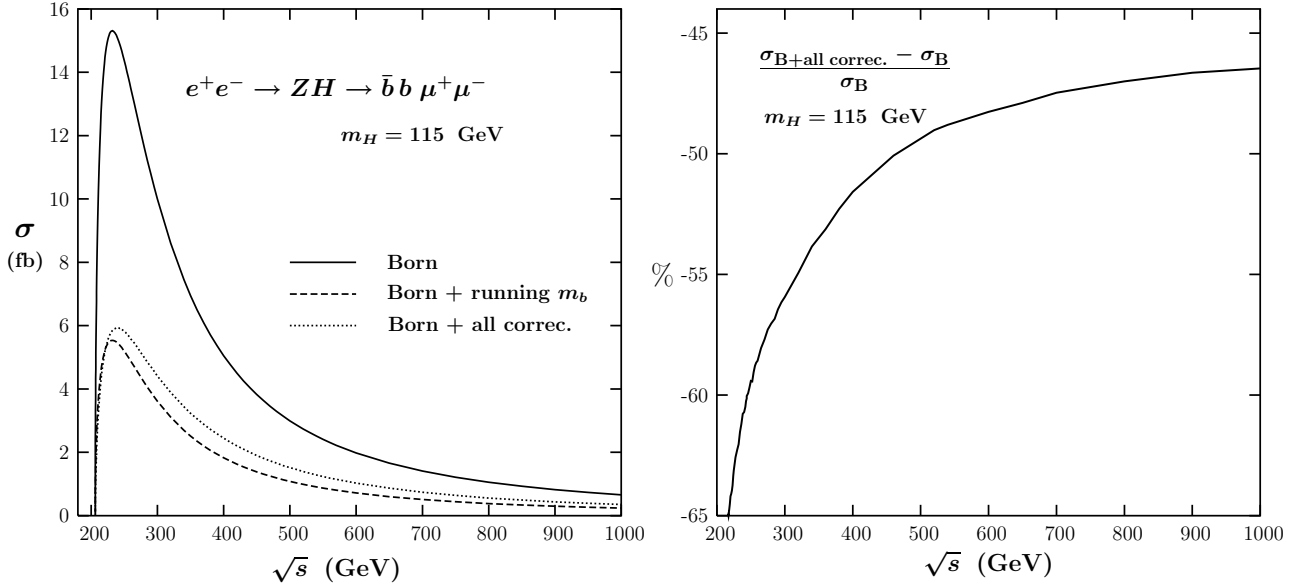

\begin{center}
\setlength{\unitlength}{1mm}
\begin{picture}(35,35)(58,-50)
\rput(5.3,-6){\scalebox{0.65 0.65}{\epsfbox{fig3a.epsi}}}
\end{picture}
\begin{picture}(35,35)(9,-50)
\rput(5.3,-6){\scalebox{0.65 0.65}{\epsfbox{fig3b.epsi}}}
\end{picture}
\end{center}
\vspace*{4.4cm}
\caption{The `signal' total cross section of reaction (\ref{bmmb}) 
         in the NWA as a function of the
         CMS energy. The plots on the left
         show the Born cross section of Eq.~(\ref{bornNWA}) (solid line), 
         the cross section of Eq.~(\ref{bornmb}) (dashed line)
         and the cross section including all corrections as given by
         Eq.~(\ref{LL}) (dotted line). The plot on the right shows 
         the full relative correction with respect to the Born cross section.}
\label{fig:sig}
\end{figure}

The total `signal' cross section of (\ref{bmmb}) in the narrow width
approximation is plotted
on the left hand side of Fig.~\ref{fig:sig} as a function of the CMS energy.
The solid curve shows the Born cross section
\bea
\label{bornNWA}
\sigma_{\rm Born}^{\rm NWA}=\sigma_{e^+e^-\ra ZH}^{(0)}\;
\frac{\Gamma_{Z\ra \mu^+\mu^-}^{(0)}}{\Gamma_Z}\;
\frac{\Gamma_{H\ra b\bar b}^{(0)}}{\Gamma_H}.
\eea
The dashed curve shows the cross section including the correction
due to running of the $b$-quark mass
\bea
\label{bornmb}
\sigma_{{\rm Born + running}\;m_b}^{\rm NWA}=\sigma_{e^+e^-\ra ZH}^{(0)}\;
\frac{\Gamma_{Z\ra \mu^+\mu^-}^{(0)}}{\Gamma_Z}\;
\frac{\Gamma_{H\ra b\bar b}^{\rm r.m.}}{\Gamma_H}.
\eea
Finally, the dotted curve shows the cross section including the complete
EW corrections to the $Z$--Higgs production and to
the $Z$ boson decay width, as well as the QCD and EW corrections to the Higgs
boson decay width obtained with {\tt HDECAY} and the ISR. The latter has
been taken into account in the structure function
approach by folding the cross section
\bea
\label{NWA}
{\rm d}\, \sigma^{\rm NWA}={\rm d}\, \sigma_{e^+e^-\ra ZH}
\frac{\Gamma_{Z\ra \mu^+\mu^-}}{\Gamma_Z}\;
\frac{\Gamma_{H\ra b\bar b}}{\Gamma_H},
\eea
with the structure function $\Gamma_{ee}^{LL}\left(x,Q^2\right)$
and integrating it over the full angular range according to
\bea
\label{LL}
\sigma_{\rm Born + all\; correc.}^{\rm NWA}=
\int_0^1 {\rm d} x_1 \int_0^1 {\rm d} x_2 \,
          \Gamma_{ee}^{LL}\left(x_1,Q^2\right)
\Gamma_{ee}^{LL}\left(x_2,Q^2\right)
  \int_{-1}^1 {\rm d}\cos\theta
\frac{{\rm d}\sigma^{\rm NWA}}{{\rm d}\cos\theta}
\left(x_1 p_1,x_2 p_2\right),
\eea
where $x_1p_1$ ($x_2p_2$) is the four momentum of the positron
(electron) after emission of a collinear photon.
The structure function $\Gamma_{ee}^{LL}\left(x,Q^2\right)$ is given
by Eq.~(67) of \cite{Beenakker}, with {\tt `BETA'} chosen for non-leading
terms. The splitting scale $Q^2$, which is not fixed in the LL approximation
is chosen to be equal $s=(p_1+p_2)^2$.
The corresponding complete relative correction is plotted
on the right hand side of Fig.~\ref{fig:sig}. It is large and dominated by
the correction to the Higgs--$b\bar b$ Yukawa coupling caused by the running
of $m_b$.

\begin{figure}[!ht]
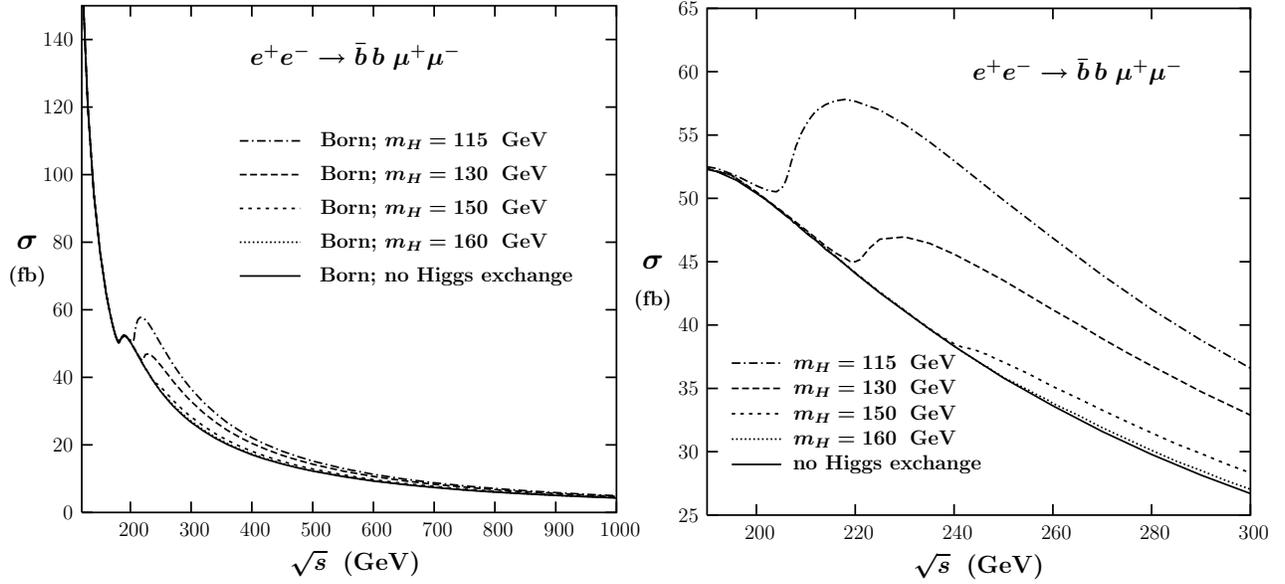

\begin{center}
\setlength{\unitlength}{1mm}
\begin{picture}(33,33)(58,-50)
\rput(5.3,-6){\scalebox{0.65 0.65}{\epsfbox{fig4a.epsi}}}
\end{picture}
\begin{picture}(35,35)(9,-50)
\rput(5.3,-6){\scalebox{0.65 0.65}{\epsfbox{fig4b.epsi}}}
\end{picture}
\end{center}
\vspace*{4.2cm}
\caption{The total Born cross section of reaction (\ref{bmmb})
          as a function of the CMS energy in the energy range
          from 120 -- 1000 GeV (left) and from 190 -- 300 GeV (right)
          for different values of $m_H$.}
\label{fig:born}
\end{figure}

How the `background' Feynman diagram contribution and off-shell effects
change the Higgsstrahlung signal cross section is illustrated
in Fig.~\ref{fig:born}, where on the left (right) hand side we plot the
total Born cross section of reaction (\ref{bmmb}) as a
function of the CMS energy in the energy range from 0.12 -- 1 TeV
(0.19 -- 0.3 TeV) for a few different values of $m_H$.
The $Z$--Higgs production signal that is clearly visible in
Fig.~\ref{fig:born} as a rise in the plots above the solid line,
which represents the cross section
of reaction (\ref{bmmb}) without the Higgs boson exchange contribution,
decreases while the Higgs mass is growing. For $m_H=150$~GeV, it becomes
already rather small, while it is hardly visible for $m_H=160$~GeV.
This effect is caused by the substantial
grow of the total Higgs boson width $\Gamma_H$ with the grow of the Higgs
boson mass $m_H$. The latter results in a decrease
of the branching ratio $\Gamma_{H\ra b\bar{b}}/\Gamma_H$, as with growing
$m_H$ the Higgs decay into $W$-boson pair starts to dominate.
The first bump in the plots on the left hand side of Fig.~\ref{fig:born}
reflects the double $Z$ production resonance.

\begin{figure}[htb]
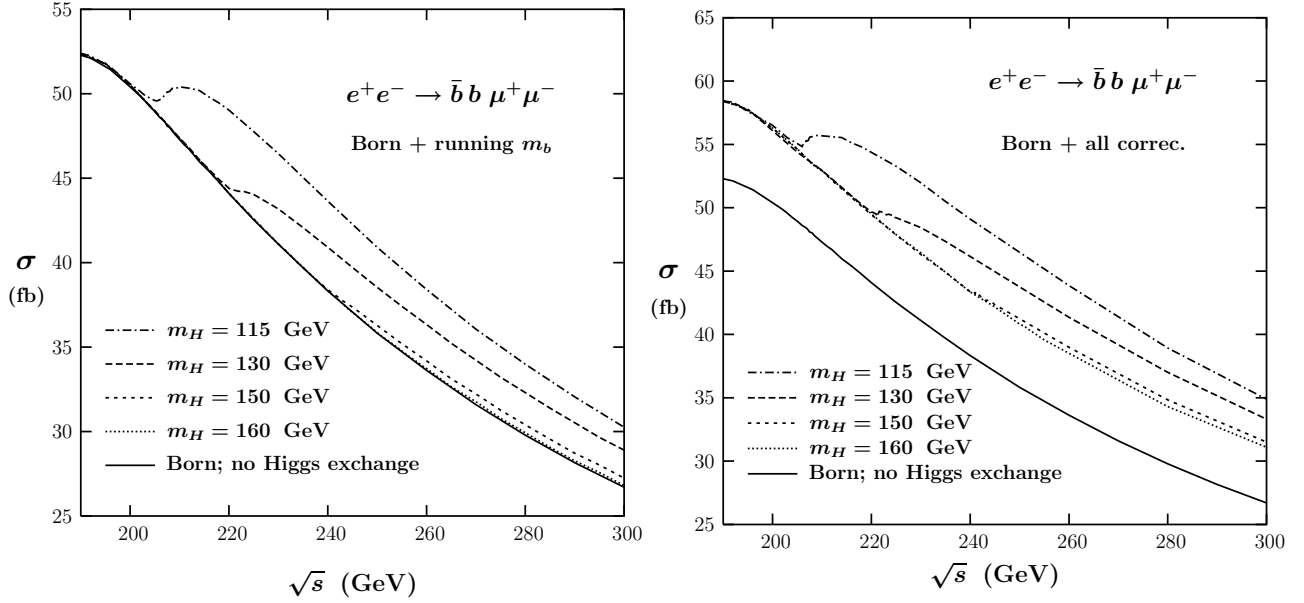

\begin{center}
\setlength{\unitlength}{1mm}
\begin{picture}(35,35)(58,-50)
\rput(5.3,-6){\scalebox{0.65 0.65}{\epsfbox{fig5a.epsi}}}
\end{picture}
\begin{picture}(33,33)(9,-50)
\rput(5.3,-6){\scalebox{0.65 0.65}{\epsfbox{fig5b.epsi}}}
\end{picture}
\end{center}
\vspace*{4.2cm}
\caption{The total cross section of reaction (\ref{bmmb})
          as a function of the CMS energy for different values of $m_H$.
          The plots on the left include
          the correction $\delta g_{Hb\bar b}^{\rm r.m.}$ to the
          Higgs--$b\bar b$ Yukawa coupling caused by the running
          of the $b$-quark mass.
          The plots on the right include all the corrections
          of Eq.~(\ref{totalcs}).}
\label{fig:tot1}
\end{figure}

\begin{figure}[!ht]
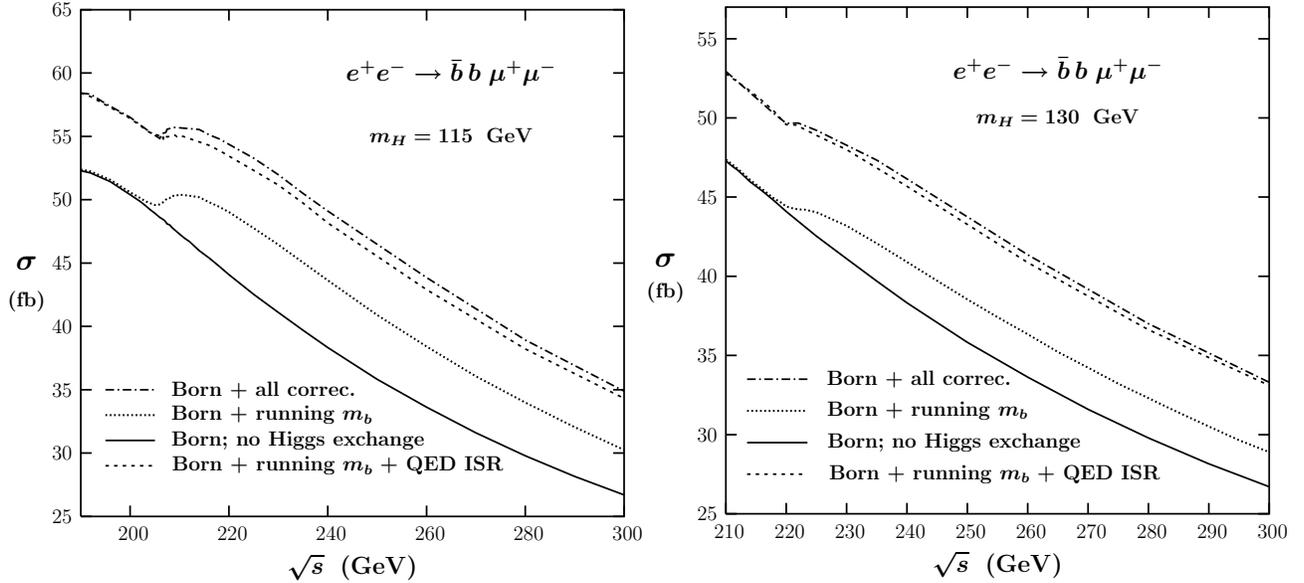

\begin{center}
\setlength{\unitlength}{1mm}
\begin{picture}(35,35)(58,-50)
\rput(5.3,-6){\scalebox{0.65 0.65}{\epsfbox{fig6a.epsi}}}
\end{picture}
\begin{picture}(33,33)(9,-50)
\rput(5.3,-6){\scalebox{0.65 0.65}{\epsfbox{fig6b.epsi}}}
\end{picture}
\end{center}
\vspace*{4.2cm}
\caption{The total cross section of reaction (\ref{bmmb}) for
         $m_H=115$~GeV (left) and $m_H=130$~GeV (right) including 
         all the corrections (dashed--dotted line) as compared to
         the cross section including the QED ISR corrections and
         the correction due to the running of $m_b$,
         corresponding to the first 3 terms on the right hand side of 
         Eq.~(\ref{totalcs}), (short dashed line). The plots of
         the Born cross section of reaction (\ref{bmmb}) without the 
         Higgs boson exchange (solid line) and the cross section of
         (\ref{bmmb}) including solely the correction due to
         the running of $m_b$ (dotted line) are shown, too.}
\label{fig:tot2}
\end{figure}

The total cross sections of reaction (\ref{bmmb}) including radiative
corrections  are plotted in Fig.~\ref{fig:tot1} as functions of the
CMS energy in the range from 190 -- 300 GeV
for a few selected values of the Higgs boson mass.
The plots on the left include the correction $\delta g_{Hb\bar b}^{\rm r.m.}$
to the Higgs--$b\bar b$ Yukawa coupling caused by the running of the
$b$-quark mass, {\em i.e.}, they depict the first integral on the right hand
side of Eq.~(\ref{totalcs}). The correction is large and it reduces
the Higgsstrahlung signal substantially as compared to the plots
on the right hand sight of Fig.~\ref{fig:born}.
The plots on the right hand side of Fig.~\ref{fig:tot1}
include all the corrections as defined in
Eq.~(\ref{totalcs}). As compared to the corresponding plots on the left,
they are shifted upwards with respect to the solid
curve representing the Born cross section of reaction (\ref{bmmb}) without
the Higgs boson contribution. This is caused by taking into account the
contribution from the initial state bremsstrahlung
with the hard photon momentum integrated over the full phase space.
As expected, inclusion of the ISR smears the $Z$--Higgs production signal.

The size of EW corrections corresponding to the fourth term on the
right hand side of Eq.~(\ref{totalcs}) can be read from Fig.~\ref{fig:tot2}, 
where the total cross section of reaction (\ref{bmmb}) for
$m_H=115$~GeV (left) and $m_H=130$~GeV (right) including 
all the corrections (dashed--dotted line) and
the cross section including the QED ISR corrections and
the correction due to the running of $m_b$,
corresponding to the first 3 terms on the right hand side of 
Eq.~(\ref{totalcs}), (short dashed line) are compared. The plots of
the Born cross section of reaction (\ref{bmmb}) without the 
Higgs boson exchange (solid line) and the cross section of
(\ref{bmmb}) including solely the correction due to
the running of $m_b$ are shown, too.

\section{SUMMARY AND OUTLOOK}

We have presented the SM predictions for a four-fermion reaction (\ref{bmmb}),
which is one of the best detection channels of a low mass Higgs boson
produced through the Higgsstrahlung mechanism at a linear collider.
We have included leading virtual and real QED ISR corrections to all the
lowest order Feynman diagrams of reaction (\ref{bmmb}). We have
modified the lowest order Higgs--$b\bar b$ Yukawa coupling in
reactions (\ref{bmmb}) and (\ref{bmmbg}) by including in it
the correction to the Higgs boson decay width caused by the running of
the $b$-quark mass, which can be calculated with a program {\tt HDECAY}.
The complete $\mathcal{O}(\alpha)$ EW corrections to the
$Z$--Higgs production and to the $Z$ decay width, as well as the
EW and remaining QCD corrections of {\tt HDECAY} have been taken
into account in the double pole approximation.
Moreover, we have illustrated how the Higgs boson production signal is visible
in the CMS dependence of the total cross section of reaction (\ref{bmmb})
for low Higgs masses and how it almost disappears for the Higgs mass
approaching the $W$-pair decay threshold.

Some further work is required in order to include
the non-factorizable corrections to reaction (\ref{bmmb}).
Although, by comparison
to a related process of the $ZZ$ production \cite{BCB}, \cite{DD}, 
one might expect that the non-factorizable corrections 
may largely cancel each other in the total cross section
and may be numerically suppressed in differential cross sections integrated
over the decay angles also for reaction (\ref{bmmb}), they should be included in
the program in order to properly take into account
radiative corrections due to final state real photon radiation 
which, in the present work, have been included inclusively
only in the EW one-loop corrections to the $Z$ and Higgs boson
decay widths.
To avoid possible negative weight events
which may occur in the MC simulation one should exponentiate the 
IR sensitive terms.


{\Large \bf Acknowledgements}

K.K. is grateful to the Alexander von Humboldt Foundation for supporting
his stay at DESY Zeuthen, where this work has been accomplished and to
the Theory Group of DESY Zeuthen for kind hospitality.

\end{document}